\documentclass[twocolumn,showpacs,preprintnumbers,amsmath,amssymb,floatfix,prl,superscriptaddress,aps]{revtex4-1}
\usepackage{blindtext}
\usepackage{graphicx}
\usepackage{float}
\usepackage[latin1]{inputenc} 
\usepackage{setspace}
\usepackage{dcolumn}
\usepackage{bm}
\usepackage{color}
\usepackage{SIunits}
\usepackage[normalem]{ulem}
\usepackage{url}
\usepackage[left, modulo, pagewise]{lineno}

\begin{document}

\preprint{PRL}

\title{Achieving extreme light intensities using relativistic plasma mirrors}
%
\author{Henri Vincenti}
\email{henri.vincenti@cea.fr}
\affiliation{%
LIDYL, CEA, CNRS, Universit\'e Paris-Saclay, CEA Saclay, 91 191 Gif-sur-Yvette, France
}%

\date{\today}

\begin{abstract}
In this letter, cutting-edge 3D Particle-In-Cell simulations are used to demonstrate that so-called relativistic plasma mirrors irradiated by PetaWatt (PW) lasers and naturally curved by laser radiation pressure can be used to tightly focus Doppler-generated harmonics to extreme intensities between $10^{25}-10^{26}W.cm^{-2}$. Those simulations are then employed to develop and validate a general 3D model of harmonic focusing by a curved relativistic plasma mirror. Finally, the insight gained from this model is used to propose novel all-optical techniques that would further increase the plasma mirror curvature with the ultimate goal of approaching the Schwinger limit. 
\end{abstract}

\pacs{Valid PACS appear here}
\maketitle

One of the main goals of Ultra-High-Intensity (UHI) physics, which investigates light-matter interactions at ultra-high light intensities, has been to constantly push forward the maximum attainable light intensities for accessing novel physical regimes. With intensities now approaching $I\approx 10^{22}W.cm^{-2}$ for PW-class lasers, UHI physics already offered remarkable opportunities to understand and model the complex laws governing plasma dynamics in the ultra-relativistic regime. 

A major challenge is now to push forward these intensities above $10^{25} W.cm^{-2}$ to access nonlinear Quantum Electrodynamics (QED) regimes barely explored so far in the lab \cite{burke1997positron,poder2018experimental}. Above this limit,  QED effects start playing a major role on the dynamics of electrons initially at rest: laser-accelerated electrons can produce $\gamma$-photons exerting a recoil comparable to electron momentum. Interaction of these $\gamma$-photons with laser photons can then produce stimulated e-/e+ pair cascades via the non-linear Breit-Wheeler process \cite{bell2008possibility,fedotov2010limitations}. Approaching $10^{29}W.cm^{-2}$ corresponding to the so-called Schwinger field $E=10^{18}V.m^{-1}$, light starts generating e-/e+ pairs out of vacuum \cite{sauter1931verhalten,Heisenberg1935,schwinger1951gauge}. Near this limit, vacuum would act as a non-linear medium whose refractive index depends on light intensity. Consequently, high intensity lasers could induce refraction of other light beams, causing breakdown of Maxwell's superposition principle that predicts that two light beams in vacuum simply add up and cannot interact with each other.  

Reaching intensities $I>10^{25}W.cm^{-2}$ should thus have a considerable impact as it would give access to a totally new type of experiments thanks to which we could validate theories of non-linear QED/extreme laser physics developed so far \cite{mourou2006optics,marklund2006nonlinear}. It would also provide insight into complex astrophysical phenomena where such non-linear QED processes occur \cite{buchanan2006thesis,ruffini2010electron}. Yet, the light intensities required to unlock those exotic regimes are more than three orders of magnitude higher than the ones delivered by current optical laser technology, hence calling for the design of novel solutions. 

One of the most promising idea to break this intensity barrier consists in inducing a Doppler frequency upshift of a laser of wavelength $\lambda$ and then focusing the up-shifted radiation of wavelength $\lambda_u\ll\lambda$ down to a focal spot size $\sigma\approx\lambda_u$. To implement this idea, a propitious path is to reflect a laser off a curved relativistic mirror. In this regard, different schemes have been proposed in the literature \cite{bulanov2003light,gordienko2005coherent}. Nevertheless, as highlighted throughout this paper, none of these has yet led to a detailed and feasible experimental proposal, mainly because they make use of idealized interaction conditions \cite{solodov2006limits} that are either not achievable or extremely difficult to control in the lab.

\begin{figure*}[ht]
    \centering
    \includegraphics[width=0.75\linewidth]{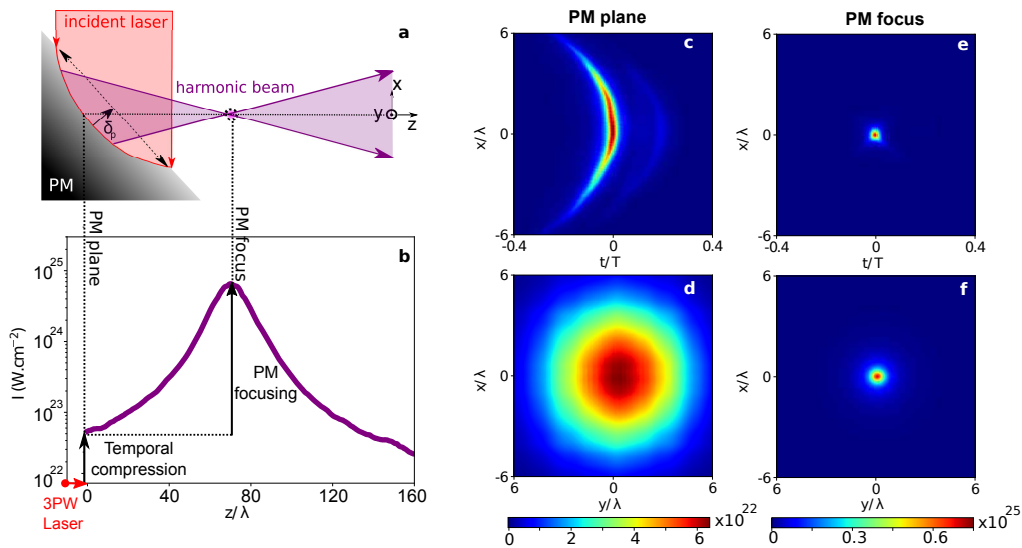}
    \caption{3D PIC simulation of PM focusing  for optimal parameters $\theta=45^o$, $L=\lambda_L/8$ and normalized laser amplitude $a_0=75$.  (a) sketch of the laser-PM interaction (b) Reflected field intensity vs distance $z$ to the PM. (c) and (e) respectively show the spatio-temporal intensity profile $I(x,t)$ of the reflected field at PM plane and PM focus. (d) and (f) respectively show the reflected beam spatial intensity profile $I(x,y)$ at PM plane and PM focus. On (c-f) the color scale represents light intensity in units of $W.cm^{-2}$.}
    \label{fig:fig1}
\end{figure*}

In this letter, I propose a novel and realistic all-optical scheme based on so-called 'plasma mirrors' \cite{kapteyn1991prepulse,doumy2004complete} (abbreviated PM) that would allow reaching intensities between $10^{25-26}W.cm^{-2}$ with PW lasers being brought into operations worldwide. The general principle of this scheme is sketched on Fig. \ref{fig:fig1} (a) and detailed below.

PMs are formed when a high power femtosecond laser with high contrast (red) is focused on an initially flat solid target. At focus, the intense laser field quasi-instantly ionizes matter and creates a dense plasma (gray scale) that specularly reflects the incident light. Upon reflection on this PM, the laser field drives relativistic oscillations of the PM surface that induce a periodic temporal compression \cite{lichters1996short,baeva2006theory,gonoskov2011ultrarelativistic} of the reflected field through the Doppler effect. This periodic modulation is associated to a high-harmonic spectrum in the frequency domain \cite{dromey2006high,thaury2007plasma,thaury2010high}. At focus, the spatially varying laser intensity (for gaussian beams) is responsible for a curvature of the PM surface \cite{dromey2009diffraction,vincenti2014optical} associated to a PM denting parameter $\delta_p$ at the center of the laser focal spot (cf. Fig. \ref{fig:fig1} (a)). This curved surface in turn focuses the high order Doppler harmonic beams (purple). As it can strongly increase harmonic divergence, this curvature effect has been considered so far as highly detrimental for applications requiring collimated harmonic beams. In this regard, several techniques have even been proposed to mitigate this effect  \cite{dromey2009diffraction,vincenti2014optical}. Instead, it is shown in this paper that the combination of (i) temporal Doppler compression and (ii) tight focusing of the Doppler harmonic beams by the radiation pressure-induced curvature is responsible for a huge light intensification by up to $3$ orders of magnitude at PM focus. 

Three-dimensional (3D) Particle-In-Cell (PIC) simulations of the interaction of a PM with at PW laser (under optimal physical conditions to be explained thereafter) were used to bring unambiguous evidence of the validity of this scheme. Such simulations are extremely challenging and could not be performed so far with standard PIC codes due to the lack of accuracy of the  finite difference Maxwell solver  \cite{blaclard2017pseudospectral,vincenti2018ultrahigh}. Thanks to the recent development and optimization of novel massively parallel and highly accurate pseudo-spectral Maxwell solvers in the PIC code WARP+PICSAR \cite{vay2013domain, vincenti2016detailed, vincenti2018ultrahigh,vay2012novel}, it was recently demonstrated that these 3D simulations can now be addressed on the largest supercomputers. In the following are presented results from a 3D PIC simulation of PMs performed with this code on the MIRA cluster at the Argonne Leadership Computer Facility (ALCF). The simulation required the full MIRA machine ($\approx 0.8$ million cores) during $24$ hours i.e a total of $\approx 20$ millions core hours. As opposed to other proposed schemes \cite{gordienko2005coherent, gonoskov2011ultrarelativistic} where such 3D 'first principles' validations are still missing, this work provides the first high-fidelity 3D PIC modelling of Doppler harmonic generation and focusing by PMs. 

The 3D simulation considered a 3PW laser of $\approx20fs$ duration with intensity $I\approx 1.2\times 10^{22}W.cm^{-2}$ (laser normalized amplitude $a_0=0.85\sqrt{I[10^{18}W.cm^{-2}]}\lambda[\mu m]\approx 75$ for a laser wavelength $\lambda=0.8\mu m$) obliquely incident with an angle $\theta=45^o$ on a PM. The laser waist is $w_L=5\lambda$. The PM  has an exponential density profile at the plasma-vacuum interface of gradient scale length $L=\lambda/8$. This can be reasonably assumed when plasma expansion, triggered by the main pulse or a controlled pre-pulse, can be considered isothermal \cite{zel2002physics}. The simulation box spans $\approx4000^3$ cells with a spatial mesh size of $\Delta\approx\lambda/200$ in all directions and a time step $\Delta t \approx T/200$ where $T$ is the laser period. 2 pseudo-particles per cell were used (see Supplemental Material sections 1-2 (SM1-2) \cite{suppmat} for detailed simulation parameters). 

Simulation results are displayed on panels (b-f) of Fig. \ref{fig:fig1}. Panel (b) shows that intensities close to $\approx 10^{25}W.cm^{-2}$ are attained at PM focus located at a position $z\approx 72\lambda$ along the specular reflection direction $z$. This intensification is first due to the periodic temporal Doppler compression of the incident laser within each laser optical cycle, just after its reflection on the relativistic oscillating PM at $z=0$. This effect leads to a factor $\approx \times 5$ intensity gain and is clearly visible on panel (c) showing the spatio-temporal intensity map of the reflected field in the PM plane over one laser optical cycle only. Besides temporal compression, the effect of PM curvature on the reflected field can be clearly observed on panel (c) showing a strong curvature of reflected field wavefronts just after reflection at $z=0$. After a propagation of the reflected field over $\approx 72\lambda$, panels (e) and (f) show a strong spatial compression of the reflected beam profile at PM focus in the transverse directions $x$ and $y$. Panel (b) shows that this additional spatial compression yields two additional orders of magnitude increase in intensity. 
\begin{figure}[ht]
    \centering
    \includegraphics[width=1\linewidth]{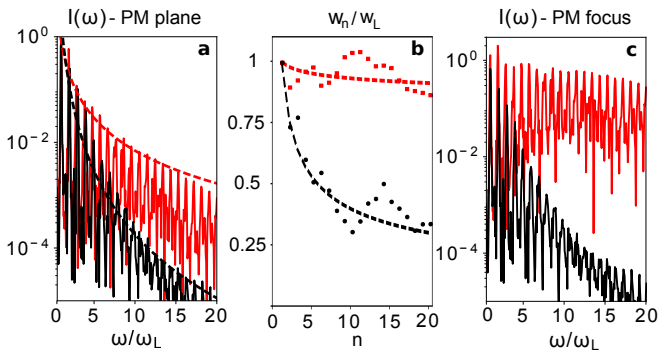}
    \caption{Effect of PM focusing on harmonic spectra. In all plots, red lines correspond to simulation results obtained for a 3PW incident laser ($a_0= 75$). The black lines correspond results obtained for a 100 TW laser case ($a_0= 2$). In all cases, $L=\lambda/8$ was used. (a) Harmonic spectra of the reflected field in the plane of the PM obtained from 2 PIC simulations. (b) Source sizes of harmonic beams obtained from 2D PIC simulations. (c) Harmonic spectra at PM focus deduced from panel (a) using equation (\ref{eq:spectrum}) and $\delta_p=\lambda/8$}
    \label{fig:fig2}
\end{figure}

One can notice that PM focusing is highly efficient and without any major optical aberrations. This can be explained by the parabolic mirror shape of the PM surface obtained with this scheme. Indeed, it was shown in \cite{vincenti2014optical} that assuming an exponential density profile of scale length $L$, the PM denting $\delta(s)$ induced by radiation pressure at position $s$ along the PM surface can be written as $\delta(s)\propto L \ln a(s)$ provided that $a(s)\gg 1$, where $a(s)$ is the spatial amplitude profile of the incident laser. For a gaussian laser beam $a(s)\propto e^{-s^2/w_L^2}$, the PM surface has thus a parabolic shape $\delta(s)\propto s^2$. Furthermore, it can be noticed that despite oblique reflection on the curved PM in the $(x,z)$ plane, no astigmatism affects the reflected field. This should indeed shorten the PM focal length $f_p$ by $\cos\theta$ in this plane compared to the $(y,z)$ plane. Yet, the oblique incidence is also responsible for the formation of an elliptical PM by radiation pressure, which has a focal $f_p$ longer by $1/\cos\theta$ in the $(x,z)$ plane. This eventually gives the same focal length in the $(x,z)$ and $(y,z)$ planes and a perfectly symmetric reflected beam at PM focus as seen on Fig. \ref{fig:fig1} (f). 

The high intensity gains obtained with the proposed scheme, via 3D simulation, are now explained quantitatively. To this end, I first derive a general model of harmonic focusing that gives the harmonic intensity gain after focusing by the PM. Results of this model are then discussed in various physical conditions using more tractable 2D PIC simulations.  

In the following, PM is assumed to have a parabolic shape with a denting $\delta_p$ at the center of the laser focal spot, as defined on Fig. \ref{fig:fig1} (a). This curved PM focuses each harmonic beam at a distance $z=z_n$, thus increasing harmonic intensity as follows: 
\begin{equation}
I_{n}^f=I_{n}^0\gamma_n^{2}
\label{eq:demagn}
\end{equation}
where $I_{n}^f$ is the harmonic intensity at $z=z_n$, $I_{n}^0$ is the harmonic intensity at $z=0$ and $\gamma_n>1$ is the demagnification factor for harmonic order $n$.  Assuming gaussian harmonics beams, the expression of $\gamma_{n}$ can be obtained as detailed in \cite{vincenti2014optical}: 
\begin{equation}
\gamma_{n}=\sqrt{1+\Psi_n^{2}}
\label{eqgam}
\end{equation}
where $\Psi_n$ is the PM dimensionless parameter defined as : 
\begin{equation}
\Psi_n=\frac{2\pi}{\cos\theta}\left(\frac{w_n}{w_L}\right)^2\frac{\delta_p}{\lambda_n}  
\end{equation}
with $\lambda_n=\lambda/n$ the harmonic wavelength and $w_n$ the harmonic source size in the PM plane. When the PM denting is much smaller than the harmonic wavelength and/or harmonic beams are generated over a too small part of the laser waist to experience the PM curvature, the PM surface does not focus harmonic beams (i.e. $\Psi_n\ll1$, $\gamma_n \approx 1$). However, in the opposite case (i.e. $\Psi_n\gg 1$, $\gamma_n\gg 1$), harmonics get focused by the PM and all harmonic orders $n$ are focused at the very same location $z_n=f_p\cos \theta$, where $f_p=w_L^2/2\delta_p$ is the focal length of the PM. In this particular case the harmonic intensity gain at PM focus writes: 
\begin{equation}
\Gamma_n=\dfrac{I_{n}^f}{I_n^0}\approx \frac{4\pi^2}{\cos^2\theta}\left(\frac{w_n}{w_L}\right)^4\left(\frac{\delta_p}{\lambda}\right)^2n^2
\label{eq:spectrum}
\end{equation}
Using equation (\ref{eq:spectrum}) and knowing the complex spectrum $E_n^0$ in PM plane (such that $I_n^0=|E_{n}^{0}|^{2}$), the total theoretical intensity gain $\Gamma$ at PM focus for the reflected field (composed of all harmonic orders) can be computed numerically from $\Gamma_n$ (see SM3 \cite{suppmat}). 

Assuming a spectrum roll-off factor $\alpha$ defined as $I_n^0=I_0/n^\alpha$, equation (\ref{eq:spectrum}) shows that one can get a harmonic intensity increasing with harmonic order $n$ at PM focus, provided that: (i) the harmonic spectrum in PM plane $I_n^0$ is slowly decaying with $n$ (i.e. $\alpha\leqslant 2$) and (ii) many harmonic orders are efficiently generated over the laser waist (i.e. $w_n/w_L\approx 1$, independent of $n$). The increase of harmonic intensity with $n$ originates from a tight focusing of high harmonic orders, initially from a source size $w_n\approx w_L$ in PM plane down to a spot size $\sigma_n\propto \lambda_n$ at PM focus, yielding large de-magnification factors $\gamma_n\approx w_L/\sigma_n\propto n$ associated to large intensity gains $\Gamma_n$ at PM focus. 
\begin{figure}[h]
    \centering
    \includegraphics[width=0.85\linewidth]{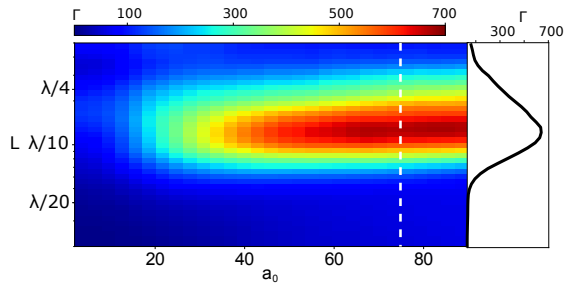}
    \caption{Intensity gain map of $\Gamma$ as a function of scale length $L$ and amplitude $a_0$ for a fixed $\theta=45^o$ obtained from 2D PIC simulations. The side panel is a line-out of the gain map along the white dashed line.}
    \label{fig:fig3}
\end{figure}

This model is now confronted to 2D PIC simulation results (cf. Fig. \ref{fig:fig2}). Red lines on panels (a)-(b) show that the optimal conditions ($\alpha\leqslant2$, $w_n/w_L\approx 1$) identified with the model are indeed met for the case of a 3PW laser and a gradient scale length $L=\lambda/8$ originally introduced in Fig. \ref{fig:fig1}. In the proposed scheme where PM curvature is induced by radiation pressure, it was shown that PM denting $\delta_p=2L\cos^2\theta$ \cite{vincenti2014optical}. For the parameters  ($\theta=45^o$,$L=\lambda/8$) used in Figs. \ref{fig:fig1} and \ref{fig:fig2}, the denting $\delta_p=L$ was $\approx0.125\lambda$ (corresponding to a radius of curvature $R=2f_p\approx 140\lambda$). Using this value, the harmonic spectrum at PM focus computed from equation (\ref{eq:spectrum}) (red line on Fig.  \ref{fig:fig2} (c)) is quasi-flat and does not vary with harmonic order. The associated $\Gamma$ computed from equation (\ref{eq:spectrum}) predicts $\approx 3$ orders of magnitude intensity gains at PM focus in perfect agreement with the results obtained from 3D PIC simulations. This intensity gain mainly comes from the focusing of harmonic orders $n$ such that $\lambda_n<\delta_p$ i.e. $n>7$ in this case.  This means that the laser itself is not focused by the PM and does not contribute to light intensification. The focusing model also indicates that only $\approx 30$ harmonic orders ($7\leqslant n\leqslant37$) contribute to $\Gamma$, which makes this scheme more robust to residual laser/PM imperfections \cite{solodov2006limits} than other schemes requiring focusing of thousands of harmonic orders \cite{gordienko2005coherent}. 

Note that the optimal conditions ($\alpha\leqslant2$, $w_n/w_L\approx 1$) are not met in the TW regime associated to lower laser intensities $\approx 10^{19}W.cm^{-2}$ (cf. black lines on Fig. \ref{fig:fig2}). In this case spectrum roll-off is much higher (cf. (a)) and harmonic source sizes are not generated efficiently over the laser waist (cf. (b)). This leads to a fast decaying spectrum at PM focus (cf. (c)) associated to an overall intensity gain $\Gamma$ that is $2$ orders of magnitude lower than the PW case. This explains why the proposed scheme expressly requires PW laser power to yield very large intensity gains.

 In the proposed scheme where the PM curvature is induced by radiation pressure, $\delta_p$ increases with $L$, which suggests that one could increase $\Gamma$ indefinitely by augmenting $L$. However, for too large values of $L$, recent studies demonstrated that harmonic efficiency can drastically decrease \cite{Kahaly2013,Chopineau2018}, therefore leading to a decrease of $\Gamma$. This suggests the existence of an optimal regime that is now determined.  
\begin{figure}[h]
    \centering
    \includegraphics[width=0.7\linewidth]{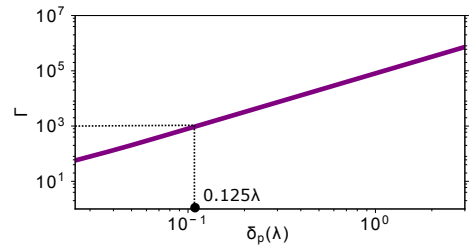}
    \caption{Evolution of $\Gamma$ with the PM denting $\delta_p$ assuming fixed parameters ($a_0=75$, $L=\lambda/8$, $\theta=45^o$) yielding constant harmonic spectra. The dashed line shows the reference case of $\delta_p=0.125\lambda$ corresponding to the case of Fig. \ref{fig:fig1} and \ref{fig:fig2}.}
    \label{fig:fig4}
\end{figure}
As there is currently no model for source sizes and spectrum roll-off (needed in the evaluation of $\Gamma_n$) as a function of laser and plasma parameters, the determination of the optimal intensity gain will entirely rely on PIC simulations. To this end, a parameter scan of $1200$ $2D$ PIC simulations was run (total of $3$ millions core hours, see SM2 \cite{suppmat} for detailed parameters), where $a_0$ was varied between $2$ and $90$ and $L$ was varied between $\lambda/50$ and $\lambda/2$ ($\theta=45^o$ was fixed). 

From this extensive set of 2D simulations, the total intensity gain $\Gamma$ at PM focus was extracted and scaled from 2D to 3D as detailed in SM4-5 \cite{suppmat}. This gain is displayed on Fig. \ref{fig:fig3}, which shows that $\Gamma$ mainly depends on $L$ for $a_0>20$. Indeed, for large enough $a_0$, harmonic beams are efficiently generated over the laser waist (i.e. $w_n/w_L\approx 1$) resulting in intensity gains $\Gamma_n$ that scale as $\delta_p^2\propto L^2$ (cf. equation \ref{eq:spectrum}). Starting at low values $L\ll \lambda$, increasing $L$ at first augments $\delta_p$, thus resulting in a rise of $\Gamma$ as seen on Fig. \ref{fig:fig3}. However, as expected, for larger values of $L$, harmonic efficiency decreases, eventually leading to a decrease of $\Gamma$. This results in the existence of an optimal value of $L \approx\lambda/8$ for $\theta=45^o$, for which $\Gamma\approx 10^3$ is maximized. These optimal parameters were precisely the ones used in the 3D simulation of Fig. \ref{fig:fig1}. This is the highest intensity gain that can be achieved by employing radiation pressure-induced curvature. 

A fascinating prospect would be to keep increasing PM curvature without degrading harmonic properties for approaching intensities close to the Schwinger limit. This could be done by finding techniques that augment $\delta_p$ independently of gradient scale length $L$, as illustrated on Fig. \ref{fig:fig4} showing the evolution of the total intensity gain $\Gamma$ with $\delta_p$ (computed using eq. (\ref{eq:spectrum})) considering fixed interaction conditions (associated to constant harmonic spectra).  As suggested by Fig. \ref{fig:fig4}, intensity gains of $\Gamma>10^5$ (i.e. intensities close to the Schwinger limit for a 3PW laser) could be achieved for $\delta_p>\lambda$ (i.e. radius of curvature $R<20\lambda$).  A realistic scheme to achieve such a control is to optically structure the initially flat solid target by pre-ionizing it with a spatially shaped pre-pulse beam. The spatial pattern of this pre-pulse beam on target modulates laser fluence, leading to a modulation of plasma temperature and expansion velocity. By the time the main laser pulse arrives, this modulated expansion velocity results in a structured plasma surface. Recent experiments have demonstrated the advanced control capabilities of this scheme \cite{Monchoce2014}, which was used to create plasma holograms \cite{Leblanc2017}. Generating a curved PM mirror with this method could be done e.g. by using a single Laguerre-Gaussian pre-pulse beam with a doughnut shape at focus. Validation of this scheme with 3D PIC simulations is in progress and will be presented in another study. Such curvature would be extremely difficult to achieve using pre-engineered solid targets in the form of $\mu m$-scale 2D/3D parabolic mirrors  \cite{gordienko2005coherent,gonoskov2011ultrarelativistic}. Such targets would indeed be extremely difficult to manufacture and align/control in experiments.

As a conclusion, this letter proposes a novel all-optical scheme to generate curved PMs from initially flat solid targets that allows for $10^3$ intensity gains at PM focus i.e. intensities of $10^{25}W.cm^{-2}$ for a 3PW laser and $10^{26}W.cm^{-2}$ for a 10PW laser. Curvature of the PM in this scheme could be controlled by properly tuning the gradient scale length $L$ \cite{vincenti2014optical}. Such control has already been demonstrated in experiments employing 100TW lasers \cite{Kahaly2013,vincenti2014optical,Chopineau2018}, which suggests that this scheme could be easily achievable on PW lasers with current experimental know-how. To go beyond the proposed scheme and approach the Schwinger limit, a favorable path would be to generate PMs with a larger curvature e.g. by employing a spatially-shaped pre-pulse beam to optically structure the PM surface. 

\begin{acknowledgements}
The author thanks F. Qu\'er\'e, G. Bonnaud and J-L Vay for precious insights during the writing of the paper. An award of computer time was provided by the INCITE program. This research used resources of the Argonne Leadership Computing Facility, which is a DOE Office of Science User Facility supported under Contract DE-AC02-06CH11357.
\end{acknowledgements}

\bibliographystyle{apsrev4-1}
\bibliography{extreme_intensity}

\end{document}